# New Insights on the High Reconnection Rate and the Diminishment of Ion Outflow

**Cheng-Yu Fan[1], Shan Wang[1*], Xu-Zhi Zhou[1], San Lu[2], Quanming Lu[2], Prayash Sharma Pyakurel[3], Qiugang Zong[1,4], Zhi-Yang Liu[5]**

[1]Institute of Space Physics and Applied Technology, Peking University, Beijing 100871, China

[2]School of Earth and Space Sciences, University of Science and Technology of China, Hefei 230026, China

[3]Space Sciences Laboratory, University of California, Berkeley, CA 94720, USA
[4] State Key Laboratory of Lunar and Planetary Sciences, Macau University of Science and Technology, Macau 999078, China
[5] Institut de Recherche en Astrophysique et Planétologie, CNES-CNRS, Universite Toulouse III, Paul Sabatier, Toulouse 31028, France

*Corresponding author: Shan Wang (coralwang90@gmail.com)

**Key Points:**

- The reconnection rate normalized by ion parameters ($R_i$) can reach high values, attributed to insufficient field line bending outside of EDR
- When $\rho_i$ exceeds the system size, the bulk velocity represents the average acceleration across the system, thus ion outflow diminishes
- $R_e$ at ~0.1 indicates a well-developed EDR; high $R_i$ indicates high absolute reconnection efficiency and weak ion coupling

## Abstract

The recently discovered electron-only reconnection has abnormal features like lack of ion outflows and high reconnection rates. Using particle-in-cell simulations, we investigate their physical mechanisms. The reconnection rate, when normalized by ion parameters ($R_i$), may appear anomalously high, whereas that normalized by electron parameters ($R_e$) remains ~0.1. We propose that the essence of high $R_i$ is insufficient field line bending outside the electron diffusion region, indicating an incomplete development of the ion diffusion region. It may result from bursty reconnection in thin current sheets, or small system sizes. The ion outflow diminishes at high $\beta_i$ when the gyroradius ($\rho_i$) exceeds the system size. Low-velocity ions still experience notable acceleration from Hall fields. However, a local distribution includes many high-velocity ions that experience random accelerations from different electric fields across $\rho_i$, resulting in near-zero bulk velocities. Our study helps understand reconnection structures and the underlying physics for transitions between different regimes.

## Plain Language Summary

Magnetic reconnection is a fundamental energy release process in plasmas. In reconnection, both ions and electrons couple to the process, receiving energizations. The reconnection rate measures the efficiency of particle acceleration and magnetic flux transfer. However, a new type of electron-only reconnection has been observed recently, where ions do not exhibit accelerations of bulk flows, yet a higher reconnection rate compared to ion-coupled reconnections is detected. We use particle-in-cell simulations to examine the controlling factors and physical meanings of the high reconnection rate and explore the reasons for the minimal ion acceleration. We find that the high reconnection rate indicates an incomplete development of ion diffusion regions with insufficient magnetic field bending, and the minimal ion bulk acceleration is due to the average over large-scale ion gyromotion covering different fields.

## 1 Introduction

Magnetic reconnection is a fundamental process in plasma physics that rearranges the topology of magnetic field lines and drives many explosive energy transferring processes in space (Birn & Priest, 2007; Yamada et al., 2010). The standard model of reconnection (Figure 1a) describes a small electron scale electron diffusion region (EDR) embedded in a large ion scale ion diffusion region (IDR). On larger scales, both ions and electrons are frozen-in with the magnetic field, but they decouple from the magnetic field within the IDR and EDR, respectively.

Recently, Phan et al. (2018) reported the electron-only reconnection in Earth's turbulent magnetosheath, where no ion-scale current sheet or ion outflow is observed throughout the reconnection process. Simulation studies provided explanations of generating electron-only reconnection by either small system sizes (e.g., smaller than 10 $d_i$, where $d_i$ is the ion inertia length) as in a turbulent environment (e.g., Pyakurel et al., 2019) or a transient early phase of standard reconnection as in studies of magnetotail current sheets (Lu et al., 2020, 2022).

The reconnection rate measures how fast the magnetic flux is reconnected, with the reconnection electric field at the X-line representing the reconnection rate per unit length of the X-line (Wang et al., 2015). It can be normalized by ion parameters ($R_i$) or electron parameters ($R_e$), both exhibiting a typical value of about 0.1 (Liu et al., 2022). However, observations show the parallel electric field $E_{\parallel}$ much larger than the expected value if normalized to ion parameters (on the order of 50) in electron-only reconnection (e.g., Phan et al., 2018; Bessho et al., 2022), which suggests a potentially large reconnection rate. Similarly, kinetic simulations have found high $R_i$ in electron-only reconnections. One of the currently accepted controlling factors is the small system size, which gives high $R_i$ and suppresses the ion outflow at the same time (Pyakurel et al., 2019). However, in simulation practices, the reconnection rate may be influenced by multiple parameters. Could any other factors also contribute to a high $R_i$? For example, the initial current sheet thickness may be a promising candidate, as demonstrated in a contemporary study by Guan et al. (2023). Furthermore, what is the essential mechanism for high $R_i$? In this study, we further examine the effects of system size and initial current sheet thickness on the reconnection rate, and try to further figure out the underlying physical meanings.

The ion dynamics in electron-only reconnection is also an intriguing problem. Phan et al. (2018) suggested that the limited space and/or time in the turbulent environments prevents ions from coupling to the magnetic structures. Simulations by Pyakurel et al. (2019) supported these findings, and they identified a system size of $10d_i$ as a critical transition point. Further research by Guan et al. (2023) revealed the transition threshold as $\rho_i$ (ion thermal gyro-radius). Despite these advances, there has yet to be a detailed investigation into the exact mechanisms to explain why ions fail to experience acceleration.

In this paper, we use particle-in-cell (PIC) simulations to explore the controlling factors and physical meanings of high reconnection rates. Additionally, we investigate the behaviour of ions in simulations with and without bulk outflows to understand why ion outflows may be failed to form.

## 2 Simulation Setup

We conducted twelve 2.5-dimensional PIC simulations described in Table 1 using the VPIC code (Bowers et al., 2008). The x boundaries are periodic, while the z boundaries are reflecting for particles and conducting for the fields. The simulations are initialized with force-free current sheets, where all initial currents are carried only by electrons. The guide field $B_g$ is equal to the upstream asymptotic field $B_{x0}$, and $B_0 = \sqrt{B_{x0}^2 + B_g^2}$. The initial magnetic field is given by $B_x = B_{x0}[tanh\,(z/L)]$ and $B_y = \sqrt{B_0^2 - B_x^2}$, where $L$ is the half thickness of the current sheet. Reconnection occurs spontaneously from numerical noises.

In Table 1, $\frac{m_i}{m_e}$ is the ion-to-electron mass ratio, $\beta_i$ ($\beta_e$) is the ion (electron) beta based on the uniform initial number density $n_0$, temperature $T_{i0}$ ($T_{e0}$) and magnetic field $B_0$. $\omega_{pe}$ is the electron plasma frequency based on $n_0$, and $\omega_{ce}$ is the electron cyclotron frequency based on $B_0$. The number of particles per cell per species (nppc) is 500 in all simulations. Unless otherwise noted, the length is normalized to the ion initial length $d_i$ (based on $n_0$), the magnetic field is in $B_{x0}$, the velocity is in $V_A = \frac{B_{x0}}{\sqrt{\mu_0 m_i}}$, and the electric field is in $B_{x0}V_A$.

The simulations are categorized into three groups. In the A and B groups, we observed the reconnection rate influenced by $L$ and $L_z$, respectively, where $L_z$ is the system size along $z$. In the C group, we altered $T_i$, and hence $\rho_i$ and $\beta_i$, to study its effect on the ion outflow.

To simulate the realistic boundary conditions, we performed three different simulation models. “1CS” refers to one initial current sheet. “3CS” represents three initial current sheets along z of the equal thickness, with the initial currents oriented in opposite $y$ directions for the adjacent current sheets. $L_z$ represents the system size along $z$ for each current sheet, so the simulation size along $z$ in the 3CS model is $3L_z$. “1+2CS” means one thin initial current sheet in the middle, with two thick initial current sheets ($L = 0.5d_i$) positioned above and below. In the 1+2CS model, $L_z$ is defined by the distance between the extreme points of $B_x$ on both sides of the thin current sheet. The thick current sheets, which do not reconnect during the time of interest, act as realistic boundaries. Thus, particles can be constrained in the z direction through the electromagnetic force instead of by artificial reflections at the boundaries. Therefore, in group C when focusing on the ion dynamics in a small system size, the 1+2CS model is applied to completely avoid the effect of artificial boundary conditions.

## 3 High reconnection rate ($R_i$) caused by insufficient magnetic field line bending

We first study the reconnection rate. Based on the diagram in Figure 1a, the normalized reconnection rates can be written as

$$R_i = \frac{\frac{dA_y}{dt}}{B_{x0}V_{Ai}}, \qquad R_e = \frac{\frac{dA_y}{dt}}{B_{xe}V_{Ae}} \tag{1}$$

Where $A_y$ is the $y$ component of the magnetic vector potential, and $V_{Ai} = \sqrt{\frac{m_e}{m_i}}V_{Ae}$ are the ion and electron Alfven speed, respectively. The simulations do not develop or have not yet developed well-defined IDRs for the intervals of interest, so we calculate $R_i$ using $B_{x0}$ at the system boundaries. $B_{xe}$ marks $B_x$ at the EDR boundaries, defined by the edge of the central $J_{ey}$ current layer. An additional $J_{ey}$ layer may develop outside the central EDR with an opposite sign, partly due to the Hall effect. Thus, we take a 1D cut of $J_{ey}$ along z at the x location of the X-line, and define the EDR boundary by considering both the maximum ($J_{ey,max}$) and minimum ($J_{ey,min}$) value of this $J_{ey}$ cut. A practical $J_{ey}$ criterion for the EDR boundary is $J_{ey} = \frac{1}{10}(J_{ey,max} + 9J_{ey,min})$, which gives reasonable and consistent results throughout the reconnection development in all runs.

With simulations in group A, we examine the effect of the initial current sheet half thickness ($L$). The reconnection rates at the time of peak $R_i$ are presented in Figure 1b. As $L$ decreases, $R_i$ increases and becomes abnormally high (approximately 0.9). Such an increasing trend of $R_i$ with decreasing $L$ is consistent with the contemporary study (Guan et al., 2023). In contrast, $R_e$ remains around 0.1. Let us compare the time history of reconnection rates in runs A1 ($L = 0.03\ d_i$, Figure 1c) and A5 ($L = 0.3\ d_i$, Figure 1d). In Figure 1c, $R_i$ (black line) rapidly reaches a high peak during the initial phase ($t\omega_{ci} = 0.2\ to\ 0.8$) before dropping to a steady phase, while $R_e$ remains consistently around 0.1 throughout the entire process. In Figure 1d, $R_i$ and $R_e$ both slowly rise to a steady value around 0.1, in accordance with the standard reconnection model.

In all of our simulations, $R_e$ remains perfectly around 0.1, so we examine the abnormally high $R_i$ by comparing it to $R_e$. The ratio between the two is

$$\frac{R_i}{R_e} = \frac{V_{Ae}B_{xe}}{V_{Ai}B_{x0}} = \sqrt{\frac{m_i}{m_e}}\left(\frac{B_{xe}}{B_{x0}}\right)^2 \tag{2}$$

This relationship shows that the high $R_i$ comes from a high $\frac{B_{xe}}{B_{x0}}$ ratio. Since $B_{xe}$ can normalize $R_e$ effectively, the high $\frac{B_{xe}}{B_{x0}}$ ratio results from a small $B_x$ reduction $\Delta B_x = B_{xe} - B_{x0}$. The magnetic field depletion outside of the EDR as illustrated in Figure 1a.

By applying Ampère's Law in the $x - z$ plane,

$$(\nabla \times B)_y = \mu_0 J_y \tag{3}$$

We take a cut at the x location of the X-line and integrate along z to obtain

$$\Delta B_x = \int_{\infty}^{EDR} (\mu_0 J_y + \frac{\partial B_z}{\partial x}) dz \tag{4}$$

The integrals $\int_{\infty}^{EDR} \mu_0 J_y dz$ (black) and $\int_{\infty}^{EDR} \frac{\partial B_z}{\partial x} dz$ (green) are presented in Figures 1e and 1f, respectively. In all of our simulations, the contribution from $\int_{\infty}^{EDR} \mu_0 J_y dz$ is much smaller than that from $\int_{\infty}^{EDR} \frac{\partial B_z}{\partial x} dz$ outside of the EDR. This phenomenon can be attributed to the fact that the $J_{ey}$ (blue) with an opposite sign from that in the central EDR, partly resulting from the Hall effect, always offsets $J_{iy}$ (red) that arises from the ion acceleration by the reconnection electric field $E_y$. Consequently, it can be approximated that

$$\Delta B_x \approx \int_{\infty}^{EDR} \frac{\partial B_z}{\partial x} dz \tag{5}$$

The term $\frac{\partial B_z}{\partial x}$ physically represents the magnetic tension, and geometrically reflects the bending of magnetic field lines in the z-direction. Thus, the low $\Delta B_x$ results from a low $\frac{\partial B_z}{\partial x}$, indicating insufficient field line bending, which further leads to high $R_i$.

To observe how a low $L$ contributes to insufficient field line bending, we examine the ratio $\frac{B_{xe}}{B_{x0}}$ (blue) in Figures 1c and 1d, which is the mean value of both sides of the current sheet. Initially, $\frac{B_{xe}}{B_{x0}}$ remains approximately 1, and then enters a rapid declining phase when the reconnection starts, followed by a transition to a steady phase which represents the quasi-steady reconnection. A key distinction between runs A1 and A5 is the evolution speed: in run A1, $R_i$ peaks within a short interval ($t\omega_{ci} = 0.5$), during the declining phase of $\frac{B_{xe}}{B_{x0}}$; in run A5, $R_i$ gradually rises to its peak ($t\omega_{ci} = 24$) during the steady phase of $\frac{B_{xe}}{B_{x0}}$. Therefore, in thin current sheets, reconnection rapidly progresses with a high reconnection rate $\frac{dA_y}{dt}$, before magnetic field lines achieve sufficient bending and hence, significant $\Delta B_x$, resulting in high $R_i$. However, $L$ does not influence the quasi-steady reconnection rate, and $R_i$ is similar in group A during the steady phase of $\frac{B_{xe}}{B_{x0}}$.

Another important factor influencing $R_i$ is $L_z$. When the field line structure extends to the system boundaries, a small $L_z$ can restrict the space available for the field lines to fully bent, resulting in higher quasi-steady $R_i$. As shown in Figure 1g for group B ($L = 0.1d_i$), both $R_i$ (black) and $\frac{B_{xe}}{B_{x0}}$ (blue) decrease as $L_z$ increases. It applies to the mean quasi-steady $R_i$ during $t\omega_{ci} = 1.6 - 3.0$. The time histories of the reconnection rate and $\frac{B_{xe}}{B_{x0}}$ for group B are provided in Figure S1, Supporting Information. After the initial declining phase, $\frac{B_{xe}}{B_{x0}}$ becomes steady, and the reconnection rate enters a quasi-steady phase. In run B3 ($L_z = 5d_i$), the restricting effect of $L_z$ is negligible, and the quasi-steady $R_i$ is similar to that in group A. Conversely, in run B1 ($L_z = 1.5d_i$), the quasi-steady $R_i$ is higher (approximately 0.4). The field line structures of runs B1 and B3 at $t_{\omega ci} = 2.5$ are presented in Figures 1i and 1j, respectively. It is evident that run B1 exhibits less field line bending compared to run B3.

To study the application and limitation of our models, we have conducted run B4 with 3CS Model to mimic turbulence environments where multiple current sheets interact. Initially, reconnection develops similarly to the 1CS model (not shown), while the later reconnection structure and timespan are significantly influenced by current sheet interactions. Figure 1h shows the $R_i$ and the movement directions of the islands in run B4 at $t\omega_{ci} = 3$. The left X-line exhibits strong $R_i$ due to the extra inflow provided by the island merging of the upper and lower current sheets. The outflow of the left X-line pushes the islands and leads to the negative $R_i$ at the right X-line. These results demonstrate the applicability of key conclusions from our simple models to turbulent conditions. However, interactions between current sheets and islands distort the sheets, potentially enhancing or reducing the lifetime and reconnection rates in later reconnection stages.

## 4 The absence of ion outflow caused by high $\beta_i$

While it is generally accepted that electron-only reconnections have high reconnection rates, it is important to note that a high reconnection rate is not necessarily associated with the absence of ion outflow. As illustrated in Figures 2a and 2c, run C1 exhibits both a high $R_i$ and a strong ion outflow.

Figures 2e-2f depict the $E_x$ and ion outflow structures in runs C1 ($\beta_i = 0.1$) and C2 ($\beta_i = 9$). $E_x$ as part of the Hall field arises from the decoupling between ions and electrons and is an important factor on accelerating ions towards the outflow (e.g., Aunai et al., 2011; Yamada et al., 2018). It is noteworthy that although the $E_x$ structures in C1 and C2 are similar, and even stronger in C2, run C1 demonstrates a strong ion outflow, while run C2 is an electron-only reconnection. A contemporary study by Guan et al. (2024) also suggests that the stronger $E_x$ structure in electron-only reconnection arises from greater charge separation due to the absence of ion outflow.

To investigate why ions in run C2 are not fully accelerated, we collected the ion velocity data from a specified area in the outflow region, indicated by the black circles in Figures 2c and 2d. We present the reduced ion velocity distribution functions (VDFs) in the $v_x - v_z$ plane of runs C1 and C2 in Figures 2g and 2h. A key difference is the ion velocity range: most ions in run C1 have a velocity not exceeding $1V_A$, whereas ions in run C2 can reach velocities of up to $10V_A$.

We examine low-velocity and high-velocity ions separately at the outflow region as marked by the circles in Figures 2c and 2d. In run C1, the low-velocity ions include all ions; in run C2, the low-velocity ions include those with velocities less than $3v_{thi,C1} = 0.9V_A$, as indicated by the white circle in Figure 2h, where $v_{thi,C1}$ is the initial ion thermal speed in C1. The one-dimensional reduced VDF along $v_x$ for only low-velocity ions is shown in Figure 2i, where both simulations exhibit a similar bulk velocity drift in the $-x$ direction. To gain further insights, we observe the typical trajectories of the low-velocity ions from both simulations in Figures 2j and 2k. These ions travel over a small scale near the X-line. Using the same color scale to represent the speed along track, they exhibit nearly identical behaviors. In conclusion, the low-velocity ions in both simulations experience accelerations and show little distinction.

For the high-velocity ions, which comprise the majority in run C2, we picture the typical trajectories in Figure 2l. One trajectory is colored in rainbow that represents the speed along track, while the other two are depicted in black and purple, respectively. These ions travel at high velocities across the system scale, as was also shown in Guan et al. (2023). The ions pass through the X-line region in a short time, which limits their acceleration from the electric fields that belong to the same X-line. Additionally, they experience counter-acceleration from the opposing electric and magnetic field throughout the system, which offsets their velocity. Consequently, it is highly possible that a single high-velocity ion cannot accumulate significant acceleration during the entire gyromotion process.

On the other hand, when considering the bulk ion outflow, we are not observing the behavior of a single particle, but rather the average velocity of all the ions within a local area. Since these local ions come from random trajectories across the $\rho_i$ scale, their average velocity will therefore represent the average acceleration across the $\rho_i$ scale. Therefore, when $\rho_i$ exceeds the system size, the bulk ion velocity reflects the average acceleration across the entire system, which is zero due to the symmetric nature of the reconnection structures.

## 5 Discussion

Our results have linked the high $R_i$ to insufficient field line bending, and it suggests that $R_i$ in such cases cannot appropriately represents the normalized reconnection rate. Comparing our study with the standard model in Figure 1a, $R_e \sim 0.1$, indicating that the reconnection structure within the EDR is already well-developed. The standard model proposes that $B_{xi}$ can properly normalize $R_i$, and in Figure 1a, we expect $B_{x0} \approx B_{xi}$. When insufficient field line bending (low $\Delta B_x$) exists between $B_{xe}$ and $B_{x0}$, $B_{x0} = B_{xe} + \Delta B_x < B_{xi}$ (the expected value of $B_{xi}$). Consequently, $B_{x0}$ cannot effectively normalize $R_i$, and $R_i$ loses its meaning as a normalized reconnection rate. The reduction of $\Delta B_x$ in small-scale reconnection was discussed in Pyakurel et al. (2019) and Bessho et al. (2022), but here we demonstrate that $\Delta B_x$ is contributed by field line bending with little contribution from the current density (e.g., discussed in Pyakurel et al., 2019). The importance of field line bending in constraining $R_i \sim 0.1$ in standard reconnection was pointed out in Liu et al. (2017). We note that the field line bending outside of the IDR would lead to $B_{x0} > B_{xi}$, but using either $B_{x0} \approx B_{xi}$ or $B_{x0} > B_{xi}$ would not alter the above analysis. Thus, our study extends the picture of field line bending to inside the IDR, concluding that insufficient bending outside of the EDR still allows for a normal reconnection efficiency within the EDR ($R_e \sim 0.1$) but fails to constrain $R_i$.

However, $R_i$ can still be meaningful as a description of the absolute value of the reconnection rate. Given a fixed boundary condition of $B_{x0}$, a higher $R_i$ means higher $\frac{dA_y}{dt} = R_i V_{Ai} B_{x0}$, i.e., more magnetic fluxes are reconnected per unit time. It serves as a useful quantity to compare the absolute reconnection rate in observation events, as long as the upstream boundary conditions can be identified.

Additionally, the high $R_i$ can be an indicator of the extent to which the ion acceleration deviates from the standard model. Since ions decouple from the field lines inside the IDR, the motion of field lines between the EDR and IDR boundaries is primarily due to the electron motion $u_{ez}$, mainly guided by the $E_y \times B_x$ drift. Therefore, insufficient field line bending can be attributed to an incompletely developed $E_y$ structure, which is a crucial factor influencing the ion acceleration within the IDR. Our simulations also show an expanding $E_y$ structure originating from the EDR, which corresponds with a similar $J_{iy}$ structure, and $u_{ix}$ grows together with $J_{iy}$ (shown in Figure S2, Supporting Information). Thus, the insufficient field line bending and the high $R_i$ are related with weak ion acceleration and incomplete IDR development. The decreasing phase of $\frac{B_{xe}}{B_{x0}}$ can represent the evolution process of IDR.

The electron-only phase can be either temporary or throughout the entire reconnection process. In all our simulations, the declining phase of $\frac{B_{xe}}{B_{x0}}$ can be defined as the temporary electron-only phase. The time development of electron and ion outflow together with $\frac{B_{xe}}{B_{x0}}$ in run A5 are exhibited in Figure S3, Supporting Information. Electron outflow starts to develop as $\frac{B_{xe}}{B_{x0}}$ starts to decline, and peaks around the end of the declining phase. Ion outflow starts to develop in the middle of the declining phase, and peaks during the steady phase of $\frac{B_{xe}}{B_{x0}}$. When the initial current sheet is thin, the reconnection rate peaks during the declining phase, producing a transient, abnormally high peak $R_i$ as in our run A1. When the initial current sheet is thick, reconnection grows slowly and the reconnection rate peaks after the declining phase, so that $R_i$ is low in the electron-only phase, such as in our run A5 and in Lu et al. (2020, 2022).In our simulation group A, with a large system size ($L_x \times L_z = 10d_i \times 5d_i$), eventually ions are coupled and $R_i$ reaches around 0.1. Thus, the incomplete ion coupling only occurs in the early time (Lu et al., 2020, 2022; Hubbert et al., 2021). In contrast, with a small system size as in our run B1 and those in Pyakurel et al. (2019, 2021), the field line bending is constrained by the system size and can never fully develop. Thus, $R_i$ remains high and ion coupling remains weak as long as reconnection proceeds with a fast $R_e \sim 0.1$.

The lack of the ion outflow is linked to high $\beta_i$. It is consistent with the decreasing trend of the ion outflow with increasing $\beta_i$ in large-scale standard reconnection (Li and Liu, 2021). Their study also found that $R_i$ decreases with higher $\beta_i$. The reconnection rates in our two simulations in group C only exhibit a small difference, so we cannot draw a conclusion on their relation. However, Guan et al. (2023) showed with multiple simulations that in small-scale reconnection, $R_i$ increases with higher $\beta_i$, opposite to the trend in large-scale reconnection. Our conjecture is that when the system size is smaller than IDR, as $\beta_i$ rises, the expected IDR size also increases. Therefore, it amplifies the weakening effect on ion acceleration due to the small system size, leading to a higher $R_i$, while $R_e$ may remain around 0.1. In large-scale reconnection, ions are magnetized once outside of the IDR, so a lower outflow at higher $\beta_i$ can be associated with a lower reconnection rate $\frac{dA_y}{dt} = E_y = V_{out}B_{out}$, where $V_{out}$ and $B_{out}$ represent the velocity and magnetic field around the outflow boundary of the IDR, respectively. Then both $R_i$ and $R_e$ would be lower. In summary, a higher $\beta_i$ leads to a weaker bulk ion outflow; it reduces the normalized reconnection rate ($R_i$ and $R_e$) in standard reconnection by really reducing the reconnection efficiency, while it can increase $R_i$ (but not $R_e$) in small-scale reconnection by introducing insufficient field line bending and incomplete IDR structures.

## 6 Conclusion

In conclusion, we have examined the normalized reconnection rates and the impact of $\beta_i$ on the ion outflow using 2.5 dimensional PIC simulations. Our findings indicate that while $R_i$ may be abnormally high, $R_e$ remains consistently around 0.1. The high $R_i$ is attributed to insufficient field line bending outside of the EDR, which corresponds to an incompletely developed IDR. A thin initial current sheet may produce temporarily high $R_i$ by fast reconnection development before field lines have been fully bent, while a small system size may limit the field line bending throughout the reconnection process. A high $R_i$ is an indicator of the extent to which ion acceleration deviates from the standard model.

Previous studies associated the small system size, high $R_i$, and lack of ion outflow together. However, we show that while high $R_i$ does represent weak ion coupling, it does not indicate a complete absence of the ion outflow. The ion outflow is suppressed by the combined effects of the small system size and high $\beta_i$ (i.e., large IDR size), rather than by the small system size alone. Hall electric fields develop in both cases in simulation group C, and low-velocity ions exhibit similar acceleration patterns. High-velocity ions in the high-$\beta_i$ run travel rapidly through the acceleration region, receiving minimal acceleration. Statistically, the local ion bulk velocity represents the average acceleration across the $\rho_i$ scale; thus, when $\rho_i$ exceeds the system scale, the ion outflow diminishes.

Our analysis advances the understanding of field structures of reconnection diffusion regions and the associated particle acceleration. As we identify the essence of $R_i$ and the ion bulk outflow, our study clarifies the relationship between normalized reconnection rates, system sizes, current sheet thicknesses, and the lack of ion outflows. The transition between different regimes related to these factors is thus understandable.

**Acknowledgements**

The study is supported by National Natural Science Foundation of China Grants 42374188. The simulation work was carried out at National Supercomputer Center in Tianjin, China, and the calculations were performed on TianHe-HPC.

**Open Research**

The simulation data presented in this paper are available (Fan et al., 2024).

**References**

Aunai, N., Belmont, G., & Smets, R. (2011). Proton acceleration in antiparallel collisionless magnetic reconnection: Kinetic mechanisms behind the fluid dynamics. *Journal of Geophysical Research Atmospheres*, *116*(A9), n/a. https://doi.org/10.1029/2011ja016688

Bessho, N., Chen, L., Stawarz, J. E., Wang, S., Hesse, M., Wilson, L. B., & Ng, J. (2022). Strong reconnection electric fields in shock-driven turbulence. *Physics of Plasmas*, *29*(4). https://doi.org/10.1063/5.0077529

Birn, J., & Priest, E. R. (2007). *Reconnection of magnetic fields : Magnetohydrodynamics and collisionless theory and observations*. Cambridge University Press.

Bowers, K. J., Albright, B. J., Bergen, B., Yin, L., Barker, K. J., & Kerbyson, D. J. (2008). 0.374 pflop/s trillion-particle kinetic modeling of laser plasma interaction on roadrunner. In *Proceedings of the 2008 ACM/IEEE Conference on Supercomputing (pp. 63:1–63:11). SC '08*. Piscataway, NJ, USA: IEEE Press.

Fan, C.-Y., Wang, S. et al. (2024). Dataset for New Insights on the High Reconnection Rate and the Diminishment of Ion Outflow. [Dataset]. Zenodo. 10.5281/zenodo.14175270

Guan, Y., Lu, Q., Lu, S., Huang, K., & Wang, R. (2023). Reconnection Rate and Transition from Ion-coupled to Electron-only Reconnection. *The Astrophysical Journal*, *958*(2), 172. https://doi.org/10.3847/1538-4357/ad05b8

Guan, Y., Lu, Q., Lu, S., Shu, Y., & Wang, R. (2024). Role of ion Dynamics in Electron-Only Magnetic Reconnection. *Geophysical Research Letters*, *51*(19). https://doi.org/10.1029/2024gl110787

Hubbert, M., Qi, Y., Russell, C. T., Burch, J. L., Giles, B. L., & Moore, T. E. (2021). Electron-Only Tail current sheets and their temporal evolution. *Geophysical Research Letters*, *48*(5). https://doi.org/10.1029/2020gl091364

Li, X., & Liu, Y. (2021). The effect of thermal pressure on collisionless magnetic reconnection rate. *The Astrophysical Journal*, *912*(2), 152. https://doi.org/10.3847/1538-4357/abf48c

Liu, Y., Cassak, P., Li, X., Hesse, M., Lin, S., & Genestreti, K. (2022). First-principles theory of the rate of magnetic reconnection in magnetospheric and solar plasmas. *Communications Physics*, *5*(1). https://doi.org/10.1038/s42005-022-00854-x

Liu, Y., Hesse, M., Guo, F., Daughton, W., Li, H., Cassak, P. A., & Shay, M. A. (2017). Why does Steady-State Magnetic Reconnection have a Maximum Local Rate of Order 0.1? *Physical Review Letters*, *118*(8). https://doi.org/10.1103/physrevlett.118.085101

Lu, S., Lu, Q., Wang, R., Pritchett, P. L., Hubbert, M., Qi, Y., Huang, K., Li, X., & Russell, C. T. (2022). Electron-Only reconnection as a transition from quiet current sheet to standard reconnection in Earth's magnetotail: Particle-In-Cell simulation and application to MMS data. *Geophysical Research Letters*, *49*(11). https://doi.org/10.1029/2022gl098547

Lu, S., Wang, R., Lu, Q., Angelopoulos, V., Nakamura, R., Artemyev, A. V., Pritchett, P. L., Liu, T. Z., Zhang, X., Baumjohann, W., Gonzalez, W., Rager, A. C., Torbert, R. B., Giles, B. L., Gershman, D. J., Russell, C. T., Strangeway, R. J., Qi, Y., Ergun, R. E., . . . Wang, S. (2020). Magnetotail reconnection onset caused by electron kinetics with a strong external driver. *Nature Communications*, *11*(1). https://doi.org/10.1038/s41467-020-18787-w

Phan, T. D., Eastwood, J. P., Shay, M. A., Drake, J. F., Sonnerup, B. U. Ö., Fujimoto, M., Cassak, P. A., Øieroset, M., Burch, J. L., Torbert, R. B., Rager, A. C., Dorelli, J. C., Gershman, D. J., Pollock, C., Pyakurel, P. S., Haggerty, C. C., Khotyaintsev, Y., Lavraud, B., Saito, Y., . . . Magnes, W. (2018). Electron magnetic reconnection without ion coupling in Earth's turbulent magnetosheath. *Nature*, *557*(7704), 202–206. https://doi.org/10.1038/s41586-018-0091-5

Pyakurel, P. S., Shay, M. A., Phan, T. D., Matthaeus, W. H., Drake, J. F., TenBarge, J. M., Haggerty, C. C., Klein, K. G., Cassak, P. A., Parashar, T. N., Swisdak, M., & Chasapis, A. (2019). Transition from ion-coupled to electron-only reconnection: Basic physics and implications for plasma turbulence. *Physics of Plasmas*, *26*(8). https://doi.org/10.1063/1.5090403

Wang, S., Kistler, L. M., Mouikis, C. G., & Petrinec, S. M. (2015). Dependence of the dayside magnetopause reconnection rate on local conditions. *Journal of Geophysical Research Space Physics*, *120*(8), 6386–6408. https://doi.org/10.1002/2015ja021524

Yamada, M., Chen, L., Yoo, J., Wang, S., Fox, W., Jara-Almonte, J., Ji, H., Daughton, W., Le, A., Burch, J., Giles, B., Hesse, M., Moore, T., & Torbert, R. (2018). The two-fluid dynamics and energetics of the asymmetric magnetic reconnection in laboratory and space plasmas. *Nature Communications*, *9*(1). https://doi.org/10.1038/s41467-018-07680-2

Yamada, M., Kulsrud, R., & Ji, H. T. (2010). Magnetic reconnection. *Reviews of Modern Physics*, **82**(1), 603–664. https://doi.org/10.1103/RevModPhys.82.603

Table 1. Plasma parameters of twelve simulation runs

<table>
<tr><th>Run</th><th>$\frac{m_i}{m_e}$</th><th>$\rho_i/d_i$</th><th>$\beta_i = \sqrt{\rho_i/d_i}$</th><th>$\beta_e$</th><th>$\frac{\omega_{pe}}{\omega_{ce}}$</th><th>$L_x/d_i$</th><th>$L_z/d_i$</th><th>$L/d_i$</th><th>$\Delta/d_i$</th><th>Model</th></tr>
<tr><td>A1</td><td rowspan="9">100</td><td rowspan="9">1.58</td><td rowspan="9">2.5</td><td rowspan="9">0.5</td><td rowspan="9">5</td><td rowspan="8">10</td><td rowspan="5">5</td><td>0.03</td><td rowspan="9">0.01</td><td rowspan="8">1CS</td></tr>
<tr><td>A2</td><td>0.05</td></tr>
<tr><td>A3</td><td>0.1</td></tr>
<tr><td>A4</td><td>0.2</td></tr>
<tr><td>A5</td><td>0.3</td></tr>
<tr><td>B1</td><td>1.5</td><td rowspan="3">0.1</td></tr>
<tr><td>B2</td><td>2</td></tr>
<tr><td>B3</td><td>5</td></tr>
<tr><td>B4</td><td>5</td><td>1</td><td>0.03</td><td>3CS</td></tr>
<tr><td>C1</td><td rowspan="2">900</td><td>0.32</td><td>0.1</td><td rowspan="2">0.2</td><td rowspan="2">2</td><td rowspan="2">1</td><td rowspan="2">1</td><td rowspan="2">0.05</td><td rowspan="2">0.001</td><td rowspan="2">1+2CS</td></tr>
<tr><td>C2</td><td>3</td><td>9</td></tr>
</table>

Note: $\frac{m_i}{m_e}$: ion-to-electron mass ratio. $\rho_i$: ion gyroradius. $\beta_i$ ($\beta_e$): ion (electron) beta . $\frac{\omega_{pe}}{\omega_{ce}}$: ratio between the electron plasma frequency and electron cyclotron frequency. $L_x$: system size along x. $L$: initial current sheet half thickness. $\Delta$: size of one cell. Model: simulation model. See text for details.

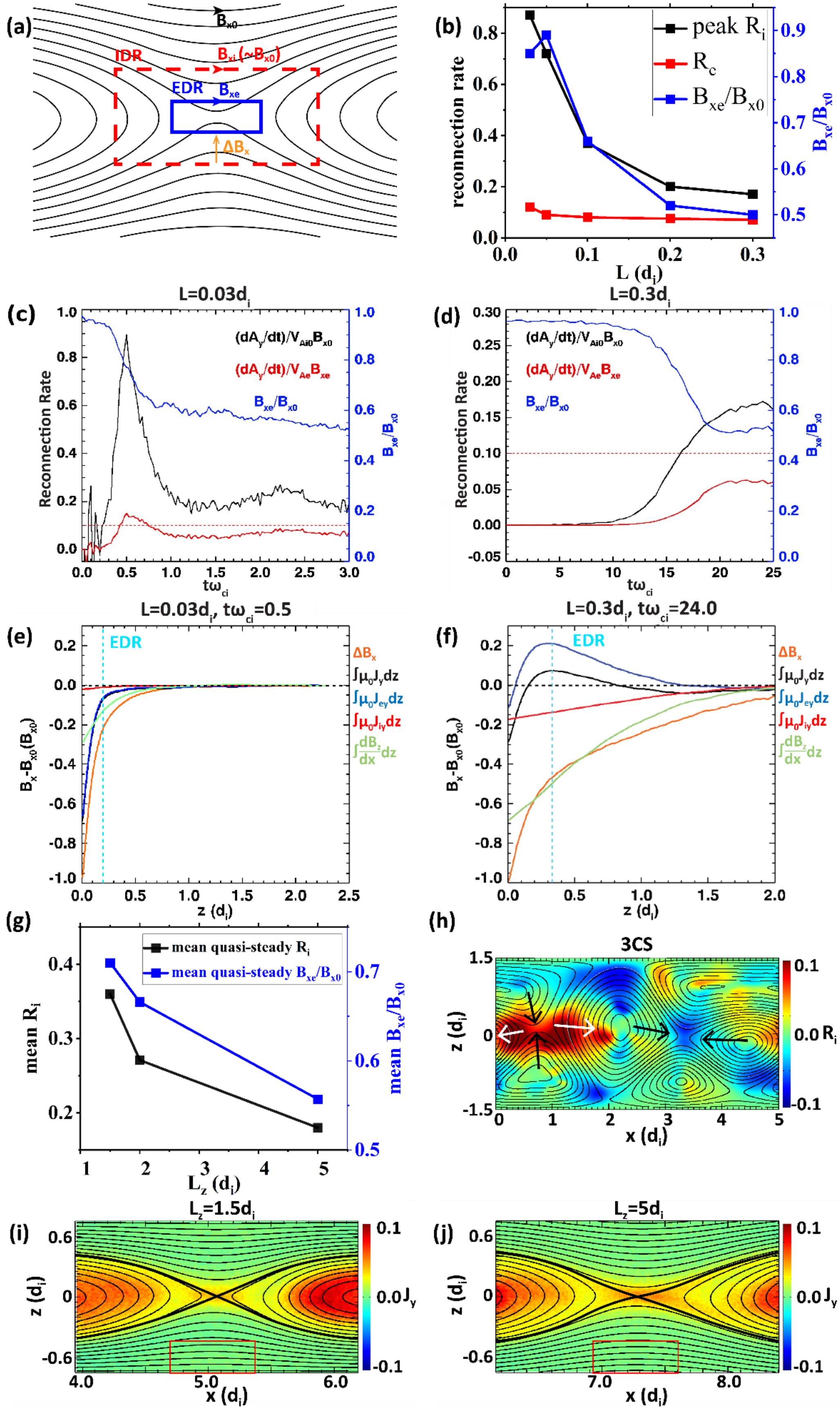

**Figure 1. (a) The structure of diffusion regions. (b) to (f): Reconnection rates affected by the initial current sheet half thicknesses ($L$). (b) Simulation Group A: Peak $R_i$, $R_e$ and $\frac{B_{xe}}{B_{x0}}$ at the time when $R_i$ reaches the peak. As $L$ increases, $R_i$ and $\frac{B_{xe}}{B_{x0}}$ both decrease, while $R_e$ always remains around 0.1. (c) and (d), runs A1 ($L = 0.03\ d_i$) and A5 ($L = 0.3\ d_i$): the evolution of $R_i$ (black), $R_e$ (red) and $\frac{B_{xe}}{B_{x0}}$ (blue) over time. In both cases, as reconnection develops, $\frac{B_{xe}}{B_{x0}}$ first decreases and then becomes stable as $R_e$ reaches around 0.1. In run A1, reconnection develops rapidly, associated with $R_i$ reaching a high peak; in run A5, $R_i$ gradually increases to around 0.1. (e) and (f), runs A1 and A5: $B_x$ (orange), the integral of $j_y$ (black) and $\frac{dB_z}{dx}$ (green) from the upper system boundary to the upper EDR boundary along the x location of the X-line at the time when $R_i$ peaks. $\Delta B_x$ is mainly contributed by $\frac{dB_z}{dx}$. Despite $j_{iy}$ (red) is larger in A5, $j_y$ remains small as $j_{iy}$ offsets $j_{ey}$ (blue). (g) Simulation group B: mean quasi-steady $R_i$ (black) and $\frac{B_{xe}}{B_{x0}}$ (blue) at $t\omega_{ci} = 1.6 - 3.0$. As $L_z$ decreases, $R_i$ and $\frac{B_{xe}}{B_{x0}}$ both decrease. (h) run B4: $R_i$ at $t\omega_{ci} = 3.0$. The reconnection structure and timespan are influenced by the current sheet interaction. The arrows represent the movement direction of the islands.(i) and (j), runs B1 ($L_z = 1.5d_i$) and B3 ($L_z = 5d_i$): the reconnection rate $R_i$ at the second $R_i$ peak. Less field line bending is observed in B1 than in B4.**

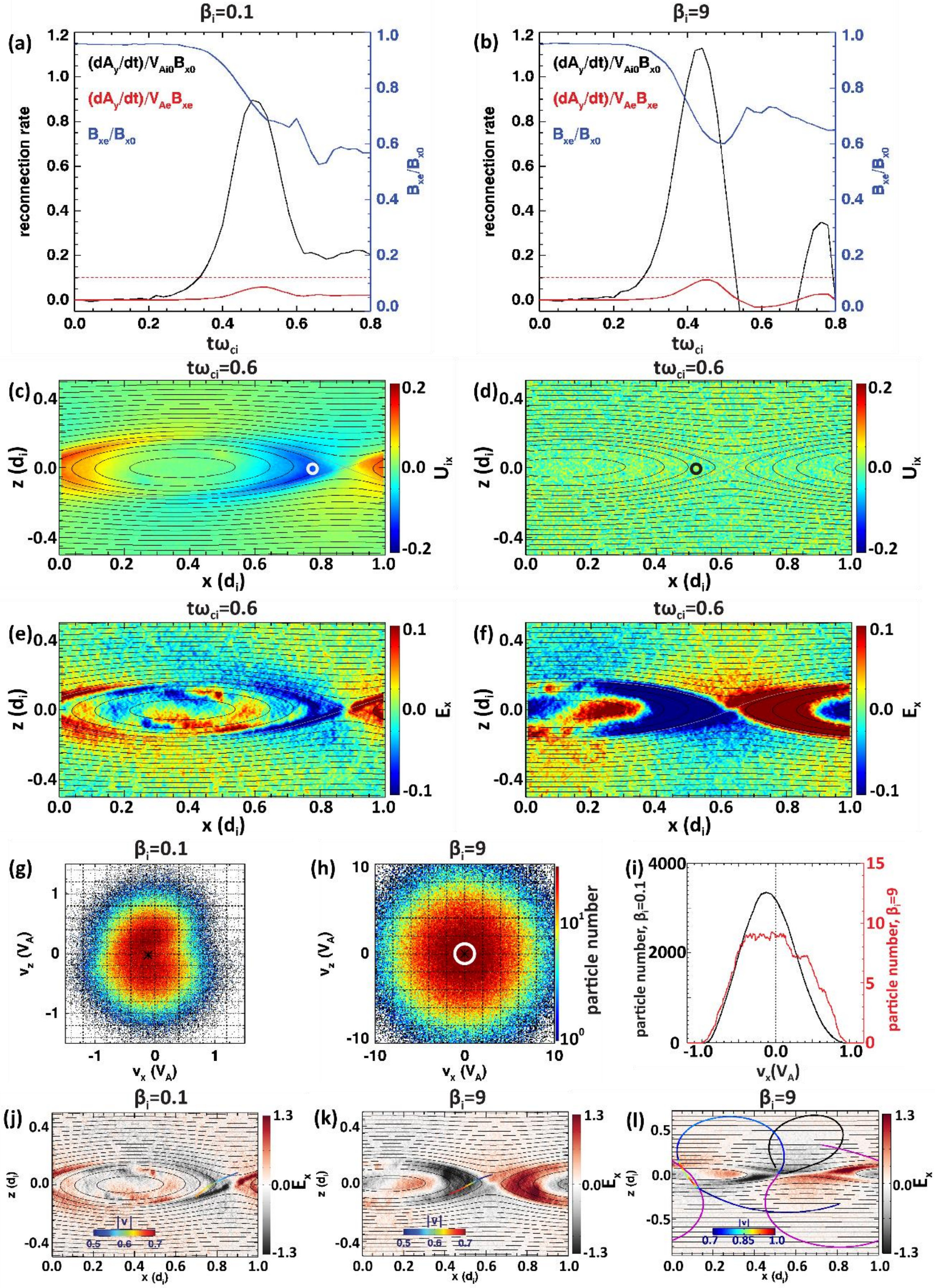

**Figure 2. (a) to (f): Reconnection ion outflows affected by $\beta_i$, run C1 (left) with $\beta_i = 0.1$ and run C2 (right) with $\beta_i = 9$. (a) and (b), the evolution of $R_i$, $R_e$ and $\frac{B_{xe}}{B_{x0}}$ over time. Both runs exhibit large peak $R_i$. (c) and (d), significant ion outflow $U_{ix}$ observed in run C1, while no ion exhaust is present in C2. White circle in (c) and black circle in (d) represent the outflow region where the ion velocity data were collected for the analysis. (e) and (f), similar $E_x$ structures observed near X-line in both runs, stronger in C2. (g) to (l): Ion particle data from the outflow region of runs C1 and C2. (g) and (h), reduced ion VDF in the $v_x - v_z$ plane. (i) One-dimensional reduced ion VDF along $v_x$. For C2 (red), only low-velocity ions with $|v| < V_A$ (within the white circle in (h)) are collected. Distributions in both runs exhibit a bulk velocity drift in the $-x$ direction. (j) and (k), typical trajectories of ions in run C1 and low-velocity ions in Run C2 overplotted on $E_x$ at $t\omega_{ci} = 0.6$. Colors along the trajectories represent their speeds, showing similar accelerations. (l) Typical trajectories of high-velocity ions in Run C2 overplotted on $E_x$ at $t\omega_{ci} = 0.4$. One of the trajectories is colored in rainbow, showing little and reversing accelerations over one cyclotron period. Two additional trajectories are shown in black and purple. Since the gyroradius exceeds the system scale, ions in the local distribution experience the acceleration across the whole system, leading to a negligible bulk velocity.**

Supporting Information for

**New Insights on the High Reconnection Rate and the Diminishment of Ion Outflow**

Cheng-Yu Fan[1], Shan Wang[1*], Xu-Zhi Zhou[1], San Lu[2], Quanming Lu[2], Prayash Sharma Pyakurel[3], Qiugang Zong[1, 4], Zhi-Yang Liu[5]

[1]Institute of Space Physics and Applied Technology, Peking University, Beijing 100871, China

[2]School of Earth and Space Sciences, University of Science and Technology of China, Hefei 230026, China

[3]Space Sciences Laboratory, University of California, Berkeley, CA 94720, USA

[4]State Key Laboratory of Lunar and Planetary Sciences, Macau University of Science and Technology, Macau 999078, China

[5]Institut de Recherche en Astrophysique et Planétologie, CNES-CNRS, Universite Toulouse III, Paul Sabatier, Toulouse 31028, France

**Contents of this file**

**Introduction**

Figure S1 includes the reconnection rates and $\frac{B_{xe}}{B_{x0}}$ in the 1CS runs in simulation group B. Figure S2 includes $R_i$, $\frac{B_{xe}}{B_{x0}}$, electron and ion outflow in runs A1 and A5. Figure S3 exhibits the development of $E_y$ and $U_{iy}$ structure in run A1.

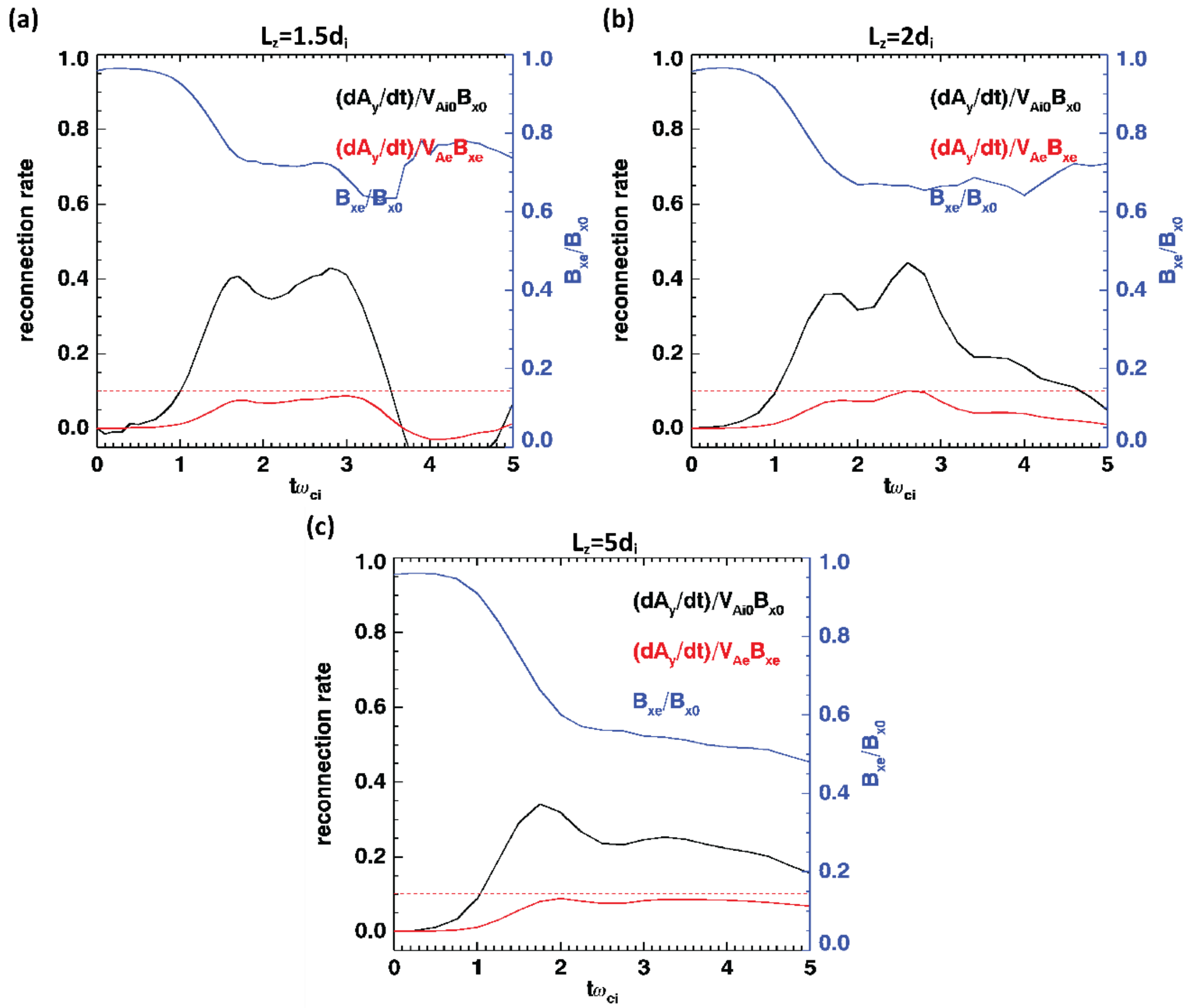

**Figure S1.** Reconnection rates affected by the system size along z ($\boldsymbol{L_z}$). (a) to (c), Runs B1 to B3: the evolution of $\boldsymbol{R_i}$ (black), $\boldsymbol{R_e}$ (red) and $\frac{B_{xe}}{B_{x0}}$ (blue) over time. The data of these time slices are collected in figure 1g.

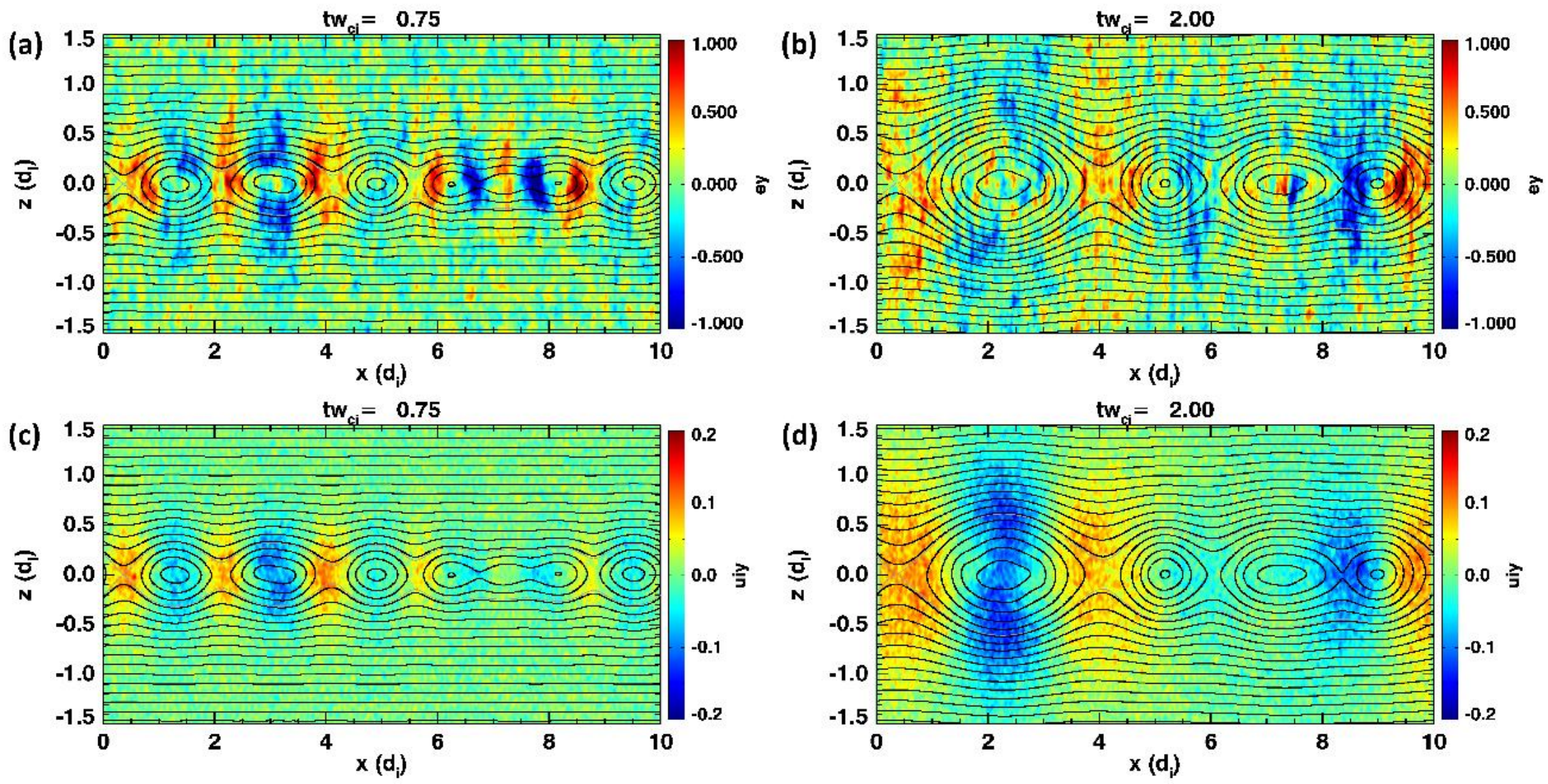

**Figure S2.** $E_y$ and $U_{iy}$ structure at $t_{\omega ci} = 0.75$ and $t_{\omega ci} = 2.0$ in run A1. Both structures expand from the EDR and develops into the IDR, representing the strengthening process of ion coupling.

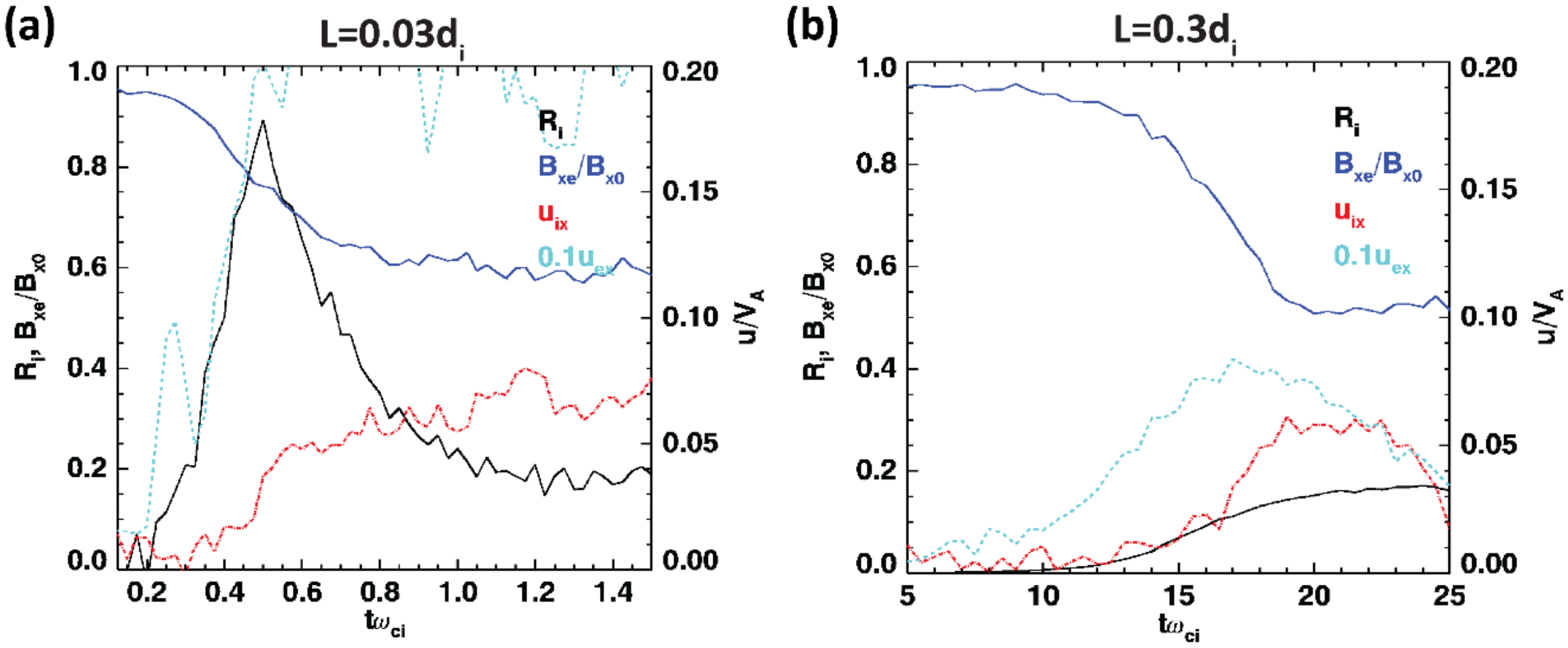

**Figure S3.** (a) and (b): Time development of $R_i$(black), $\frac{B_{xe}}{B_{x0}}$ (blue), electron (blue dashed) and ion (red dashed) outflow. Electron outflow starts to develop as $\frac{B_{xe}}{B_{x0}}$ starts to decline, and peaks around the end of the declining phase. Ion outflow starts to develop in the middle of the declining phase, and peaks during the steady phase of $\frac{B_{xe}}{B_{x0}}$.